\documentclass[12pt,a4paper]{article}

\usepackage[nosort]{cite}
\usepackage[hyperref,bulletsep]{collect}
\usepackage{graphicx}
\usepackage{pst-all}
\usepackage{bbm}

\makeatletter \@addtoreset{equation}{section} \makeatother

\makeatletter
\let\old@startsection=\@startsection
\let\oldl@section=\l@section
\renewcommand{\@startsection}[6]{\old@startsection{#1}{#2}{#3}{#4}{#5}{#6\mathversion{bold}}}
\renewcommand{\l@section}[2]{\oldl@section{\mathversion{bold}#1}{#2}}
\makeatother

\makeatletter
\let\old@makecaption=\@makecaption
\def\@makecaption{\small\old@makecaption}
\makeatother

\usepackage{amsmath,amssymb,graphicx}

\def\appendix#1{
  \addtocounter{section}{1}
 \setcounter{equation}{0}
  \renewcommand{\thesection}{\Alph{section}}
 \section*{Appendix \thesection\protect\indent \parbox[t]{11.715cm} {#1}}
  \addcontentsline{toc}{section}{Appendix \thesection\ \ \ #1}
  }

\renewcommand{\thefootnote}{\arabic{footnote}}
\let\oldPhi=\Phi
\let\oldPsi=\Psi
\let\oldGamma=\Gamma
\let\oldDelta=\Delta
\let\oldSigma=\Sigma
\let\oldTheta=\Theta
\let\oldPi=\Pi
\let\oldUpsilon=\Upsilon
\renewcommand{\Phi}{\mathnormal{\oldPhi}}
\renewcommand{\Psi}{\mathnormal{\oldPsi}}
\renewcommand{\Gamma}{\mathnormal{\oldGamma}}
\renewcommand{\Sigma}{\mathnormal{\oldSigma}}
\renewcommand{\Delta}{\mathnormal{\oldDelta}}
\renewcommand{\Theta}{\mathnormal{\oldTheta}}
\renewcommand{\Pi}{\mathnormal{\oldPi}}
\renewcommand{\Upsilon}{\mathnormal{\oldUpsilon}}

\renewcommand{\Re}{\mathop{\mathrm{Re}}}
\renewcommand{\Im}{\mathop{\mathrm{Im}}}

\ifx\genfrac\sdflkaj

\else

\fi

\newcommand{\ket}[1]{\mathopen{|}#1\mathclose{\rangle}}

\newcommand{\bigabs}[1]{\bigl|#1\bigr|}

\def\[{\begin{equation}}
\def\]{\end{equation}}

\makeatletter
\def\mr@ignsp#1 {\ifx\:#1\@empty\else #1\expandafter\mr@ignsp\fi}%
\newcommand{\multiref}[1]{\begingroup
\xdef\mr@no@sparg{\expandafter\mr@ignsp#1 \: }%
\def\mr@comma{}%
\@for\mr@refs:=\mr@no@sparg\do{\mr@comma\def\mr@comma{,}\ref{\mr@refs}}%
\endgroup}
\makeatother

\newcommand{\hypref}[2]{\ifx\href\asklfhas #2\else\href{#1}{#2}\fi}

\renewcommand{\eqref}[1]{(\multiref{#1})}

\ifx\href\asklfhas\newcommand{\href}[2]{#2}\fi

\newcommand{\comma}{\quad,\quad}
\newcommand{\unit}{\mathbbm{1}}

\newcommand{\g}{\gamma}

\newcommand{\Smatrix}{\mathbbm{S}}  

\newcommand{\be}{\begin{eqnarray}}
\newcommand{\ee}{\end{eqnarray}}

\begin{document}

\thispagestyle{empty}
\begin{flushright}\footnotesize
\texttt{hep-th/0701240}\\
\texttt{ITEP-TH-06/07}\\ \texttt{UUITP-02/07} \vspace{0.8cm}
\end{flushright}

\renewcommand{\thefootnote}{\fnsymbol{footnote}}
\setcounter{footnote}{0}

\begin{center}
{\Large\textbf{\mathversion{bold} Reduced sigma-model on
$AdS_5\times S^5$: \\
one-loop  scattering amplitudes}\par}

\vspace{1.5cm}

\textrm{T.~Klose and K.~Zarembo\footnote{Also at ITEP, Moscow,
Russia}} \vspace{8mm}

\textit{
Department of Theoretical Physics, Uppsala University\\
SE-751 08 Uppsala, Sweden}\\
\texttt{Thomas.Klose,Konstantin.Zarembo@teorfys.uu.se} \vspace{3mm}


\par\vspace{1cm}

\textbf{Abstract} \vspace{5mm}

\begin{minipage}{14cm}
We compute the one-loop S-matrix in the reduced sigma-model which
describes $AdS_5\times S^5$ string theory in the near-flat-space
limit. The result agrees with the corresponding limit of the
S-matrix in the full sigma-model, which demonstrates the consistency
of the reduction at the quantum level.
\end{minipage}

\end{center}

\vspace{0.5cm}

\renewcommand{\thefootnote}{\arabic{footnote}}
\setcounter{footnote}{0}



\section{Introduction}

According to the AdS/CFT correspondence, the large-$N$ string of the
super-Yang-Mills (SYM) theory in four dimensions has $AdS_5\times
S^5$ as the target space
\cite{Maldacena:1998re,Gubser:1998bc,Witten:1998qj}. The sigma-model
on $AdS_5\times S^5$ \cite{Metsaev:1998it} is a complicated
interacting field theory whose direct solution is currently beyond
reach. A relatively simple kinematic truncation of the AdS
sigma-model was proposed recently by Maldacena and Swanson
\cite{Maldacena:2006rv}. Technical simplifications brought about by this trunctation
potentially allow one to test various guesses and conjectures
about the AdS/CFT correspondence, and eventually can help in
quantizing strings in $AdS_5\times S^5$. The purpose of this paper
is to test the quantum mechanical consistency of this truncation.

The left and right movers on the world-sheet of the AdS string mix
and cannot be factored from each other. Maldacena and
Swanson proposed to separate what is as close to the left-moving
sector as it could be, namely to consider modes whose right-moving
momentum $p_+$ is much smaller than $p_-$. Massive string states in
the near-flat-space limit of $AdS_5\times S^5$ \cite{Gubser:1998bc}
are built precisely from these asymmetric modes \cite{Arutyunov:2004vx}. As was
shown in \cite{Maldacena:2006rv}, the action of the sigma-model can
be consistently truncated and considerably simplified, if $p_\pm$
scale as $p_\pm\sim \lambda ^{\mp 1/4}$, where $\lambda $ is the
(large) 't~Hooft coupling of SYM and $2\pi /\sqrt{\lambda }$ is
the (small) loop-counting parameter of the sigma-model. However, it
is not at all obvious if such truncation is quantum mechanically
consistent. Keeping only high-energy modes in the external legs does not
guarantee that low-momentum modes do not appear as intermediate
states in quantum loops.

To test the quantum consistency of the truncation we will calculate
the one-loop S-matrix in the reduced model and compare it to the
corresponding limit of the complete S-matrix. The S-matrix plays an
important role in the AdS/CFT correspondence
\cite{Staudacher:2004tk} because both planar SYM
\cite{Minahan:2002ve,Beisert:2003tq,Beisert:2003yb} and the string
sigma-model \cite{Bena:2003wd,Kazakov:2004qf} are completely
integrable, and their common spectrum is in principle completely
determined by feeding the two-body S-matrix in the Bethe equations
\cite{Beisert:2005fw}. The tensor structure of the two-particle
S-matrix is determined by symmetries
\cite{Beisert:2005tm}\footnote{The scattering matrix of the string
modes \cite{Klose:2006zd} differs from the gauge-theory S-matrix
\cite{Beisert:2005tm} by a scattering-state dependent transformation
that brings the S-matrix to the canonical form
\cite{Arutyunov:2006yd}.}. The left over freedom is given by an
abelian phase factor which has not been derived from first
principles but may be already known exactly: the phase of the
S-matrix satisfies a linear functional equation as a consequence of
the crossing symmetry \cite{Janik:2006dc}; the tree-level phase can
be extracted \cite{Arutyunov:2004vx} from classical Bethe equations
\cite{Kazakov:2004qf} or from the scattering of giant magnons
\cite{Hofman:2006xt}; the one-loop phase was guessed in
\cite{Hernandez:2006tk} and is well tested by comparing one-loop
corrections to various classical string configurations
\cite{Frolov:2002av,Frolov:2003tu,Frolov:2004bh,Park:2005ji,Frolov:2006qe}
to the Bethe-ansatz predictions
\cite{Schafer-Nameki:2005tn,Beisert:2005cw,Schafer-Nameki:2005is,Schafer-Nameki:2006gk,Freyhult:2006vr,Schafer-Nameki:2006ey,Benna:2006nd};
an all-order solution of the crossing equation was found in
\cite{Beisert:2006ib} and the non-perturbative phase factor valid
for the whole range of $\lambda $ was proposed in
\cite{Beisert:2006ez}. We will see that the one-loop amplitude in
the truncated sigma-model perfectly agrees with the one-loop phase
in \cite{Hernandez:2006tk}. This is not so much a check of the
latter, but rather a check of the quantum consistency of the
near-flat space limit. The agreement means that the low momentum
modes which are projected out in the reduced theory do not show up
in the one-loop amplitudes or that their contribution cancels for
external states with large $p_-$. In other words, the near-flat
space limit and the one-loop computation commute.

\section{Reduced sigma-model}

The action of the reduced model in the light-cone gauge is
\cite{Maldacena:2006rv}:
\begin{eqnarray}\label{1}
 \frac{1}{4}\mathcal{L}&=&
 \partial _+Y\partial _-Y-\frac{1}{4}\,Y^2
 +\partial _+Z\partial _-Z-\frac{1}{4}\,Z^2
 +i\,\psi _+\partial _-\psi _+
 +i\,\psi _-\Pi \psi _+
 \nonumber \\
 &&+i\,\psi _-\partial _+\psi _- +\left(Y^2-Z^2\right)\left[\left(\partial _-Y\right)^2+
 \left(\partial _-Z\right)^2\right]
 +i\left(Y^2-Z^2\right)\psi _-\partial _-\psi _-
 \nonumber \\
 &&
 +i\psi _-\left(\partial _-Y^{i'} \Gamma^{i'}
 +\partial_-Z^i\Gamma ^i \right)
 \left(Y^{i'} \Gamma^{i'}-Z^i\Gamma ^i \right)\psi _-
 \nonumber \\
 &&
 -\frac{1}{24}\left(\psi _-\Gamma ^{i'j'}\psi _-\,\psi _-\Gamma ^{i'j'}\psi _-
 -\psi _-\Gamma ^{ij}\psi _-\,\psi _-\Gamma ^{ij}\psi _-
 \right) \; .
\end{eqnarray}
The four-component bosonic fields $Y^{i'}$ and $Z^i$ describe string
fluctuations in the $S^5$ and $AdS_5$ directions, respectively. The
fermionic fields $\psi _\pm$ are eight-dimensional Majorana-Weyl
spinors of the same chirality, $\Gamma ^I$ are real $SO(8)$ Dirac
matrices and $\Pi =\Gamma ^1\Gamma ^2\Gamma ^3\Gamma ^4$. $\partial
_\pm$ are the usual light-cone derivatives: $\partial _\pm=(\partial
_0-\partial _1)/2$. The Lagrangian (\ref{1}) does not depend on any
parameters, dimensionful or dimensionless, but for making the
power-counting easier it is convenient to introduce such parameters
by rescaling the world-sheet coordinates and the fermions as $\sigma
^\pm\rightarrow \g^{\pm 1/2}m\sigma ^\pm$, $\psi _\pm\rightarrow
\g^{\mp 1/4}m^{-1/2}\psi _\pm$ (this is a combination of a
dilatation and a boost). It is also convenient to rescale all the
fields by a factor of $1/\sqrt{2}$ which brings the kinetic terms to
the canonical form. In addition we integrate out $\psi _+$ which
enters the action quadratically. After all these transformations and
dropping the minus from $\psi_-$, the
action becomes
\begin{eqnarray}\label{main}
 \mathcal{L}&=&\frac{1}{2}\left(\partial
 Y\right)^2-\frac{m^2}{2}\,Y^2
 +\frac{1}{2}\left(\partial Z\right)^2-\frac{m^2}{2}\,Z^2
 +\frac{i}{2}\,\psi \,\frac{\partial ^2+m^2}{\partial _-}\,\psi
 \nonumber \\
 &&
 +\g\left(Y^2-Z^2\right)\left[\left(\partial _-Y\right)^2
 +\left(\partial _-Z\right)^2\right]
 +i\g\left(Y^2-Z^2\right)\psi \partial _-\psi
 \nonumber \\
 &&
 +i\g\psi\left(\partial _-Y^{i'} \Gamma^{i'}
 +\partial_-Z^i\Gamma ^i \right)
 \left(Y^{i'} \Gamma^{i'}-Z^i\Gamma ^i \right)\psi
 \nonumber \\
 &&
 -\frac{\g}{24}\left(\psi\Gamma ^{i'j'}\psi\,\psi\Gamma ^{i'j'}\psi
 -\psi\Gamma ^{ij}\psi\,\psi\Gamma ^{ij}\psi
 \right)\;.
\end{eqnarray}
The truncated model (\ref{1}) was obtained from the sigma-model on
$AdS_5\times S^5$ by a boost in the $\sigma ^-$ direction with
rapidity $\sim \lambda ^{1/4}\gg 1$ \cite{Maldacena:2006rv}. The
rescaling with $\g$ essentially undoes the boost and by setting
\begin{equation}\label{glambda}
 \g=\frac{\pi }{\sqrt{\lambda }} \; ,
\end{equation}
we make the kinemtical variables in the truncated model the same as
in the original sigma-model, assuming that $p_-\gg p_+$ in the
latter. The numerical coefficient in (\ref{glambda}) is most easily
fixed by comparing the tree-level scattering amplitudes with those
in the sigma-model \cite{Klose:2006zd}. The mass $m$ should be set
to $1$ at the end of the calculation.

\begin{figure}[t]
\centerline{\includegraphics[width=8cm]{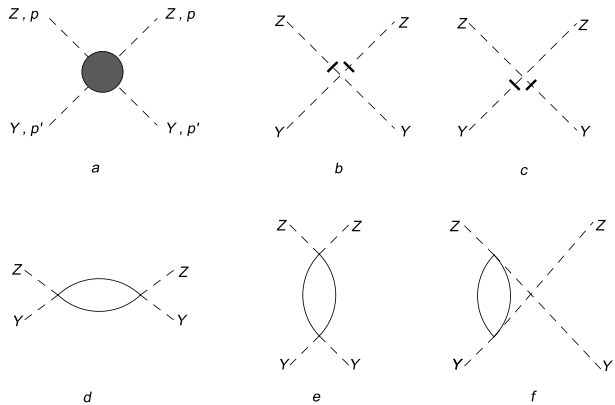}}
\caption{\label{scatter}\small  The scattering amplitude
$ZY\rightarrow ZY$. Diagrams (b) and (c) contribute at tree level.
The one-loop amplitude contains s-, t-, and u-channel diagram: (d),
(e) and (f). Any of the fields $Z$, $Y$ and $\psi $ can propagate in
the loop. }
\end{figure}

We now turn to the calculation of the S-matrix. Since all its
components are related by symmetry, it suffices to calculate only
one matrix element. We will compute the forward $ZY\rightarrow ZY$
scattering amplitude, drawn in fig. \ref{scatter}~(a).  This particular
amplitude is chosen because its calculation involves the least
amount of combinatorics, which becomes rather cumbersome already at
the one-loop level.

\subsection{Tree-level amplitude}

 In two dimensions
$2\rightarrow 2$ scattering has no phase space, and particles can
either preserve or exchange their momenta, since the conservation
condition of the two-momentum can be written as
\begin{equation}\label{momcons}
 \delta (p+p'-k-k')
 =\frac{p_0p'_0}{p'_0p_1-p_0p'_1}\,
 \Bigl(\delta (p_1-k_1)\delta (p_1-k'_1)+\delta (p_1-k'_1)
 \delta (p'_1-k_1)\Bigr) \; .
\end{equation}
We  consider the transition amplitude $Z(p)Y(p')\rightarrow
Z(p)Y(p')$ which amounts in keeping only the first delta-function in
the right-hand side. The Jacobian in (\ref{momcons}) and
relativistic normalization factors in the wave functions
($1/\sqrt{2p_0}$ for each external line) combine into an extra
factor
\begin{equation} \label{jacobian}
 \frac{1}{4}\,\frac{1}{p'_0p_1-p_0p'_1}=
 \frac{p_-p'_-}{2m^2\left(p'{}^2_--p_-^2\right)}
\end{equation}
that should be taken into account when extracting the S-matrix
elements from Feynman diagrams. Here $p_\pm=(p_0 \pm p_1)/2$.

At tree level we need to evaluate just two diagrams, (b) and (c) in
fig.~\ref{scatter}. A simple calculation gives\footnote{We put $m=1$
here.}:
\begin{equation}\label{}
 S = 1 - 2i\g p_-p'_- + \mathcal{O}(\g^2) \; .
\end{equation}
Upon identification (\ref{glambda}), this agrees with the tree-level
scattering amplitude in the sigma-model \cite{Klose:2006zd} in the
limit
\[ \label{nearflatlimit}
  p_-\rightarrow \infty \comma p_+p_-=\mbox{fixed} \; .
\]

\subsection{One-loop amplitude}

The one-loop diagrams are shown on fig.~\ref{scatter}~(d), (e), (f). There
are also several ways of distributing the derivatives in the vertices
among various lines. The superficial degree of divergence of these
diagrams is zero, which potentially leaves room for logarithmic UV
divergences. Nevertheless, all the diagrams turn out to be finite.
There are two reasons for that. First, fermi-bose cancelations
reduce the degree of divergence by one and, second, the integrands
behave as $k_-^2/k^4$ at large momenta, which gives zero upon
angular integration even before the cancelations are taken into
account.

Using Feynman rules that follow from the Lagrangian (\ref{main}), we
find for the one-loop amplitude (which has to be divided by the
Jacobian \eqref{jacobian} to get the S-matrix element):
\begin{eqnarray}\label{}
 \mathcal{A}_{\rm 1-loop}&=&16\g^2
 \left(p_-^2+p'^{2}_-\right)\left\{
 \int_{}^{}\frac{d^2k}{\left(2\pi \right)^2}\,
 \frac{\left(p_-+p'_-\right)k_-}{\left(k^2-m^2\right)
 \left[\left(p+p'-k\right)^2-m^2\right]}
  \right.\nonumber \\
  &&\qquad\qquad\qquad\;\;+\left.
 \int_{}^{}\frac{d^2k}{\left(2\pi \right)^2}\,
 \frac{\left(p_--p'_-\right)k_-}{\left(k^2-m^2\right)
 \left[\left(p-p'-k\right)^2-m^2\right]}
 \right\}
 \nonumber \\ &&+64\g^2p_-^2 p'{}^2_-
 \int_{}^{}\frac{d^2k}{\left(2\pi \right)^2}\,
 \frac{1}{\left(k^2-m^2\right)^2} \; .
\end{eqnarray}
The first two integrals correspond to the two-particle exchange in
the s- and u-channels. The last term is the t-channel contribution.
The s-channel amplitude contains an absorptive part from the
on-shell intermediate states. This can be related to the tree
amplitudes by unitarity (see below). The u-channel diagram is an
analytic continuation of the s-channel contribution to Euclidean
momenta. This happens to lead to additional cancelations, and the
final result takes a  relatively simple form\footnote{Again we set
$m=1$.}:
\begin{equation}\label{S-1loop}
 S_{\rm 1-loop}=\frac{8i\g^2p_-^3p'{}^3_-}{\pi \left(p'{}^2_--p_-^2\right)}
 \left(1-\frac{p'{}^2_-+p_-^2}{p'{}^2_--p_-^2}\,\ln\frac{p'_-}{p_-}\right)
 -
 \frac{2\g^2p_-^2p'{}^2_-\left(p'{}^2_-+p_-^2\right)}
 {\left(p'_--p_-\right)^2} \; .
\end{equation}

\section{Near-flat space limit of canonical S-matrix}
\label{sec:nearflat}

In this section we compare our one-loop results to the
$p_-\rightarrow \infty $ limit of the S-matrix of the full string
sigma-model. The exact S-matrix is expressed in terms of the
following kinemtical variables\footnote{Our normalization of the momenta
is different by a factor of $2\pi /\sqrt{\lambda }$ from the one
commonly used in the literature. This normalization is natural from
the point of view of the perturbative sigma-model
\cite{Klose:2006zd}.}
\begin{equation}\label{xpm}
x_{\pm}(p) = \frac{1+\sqrt{1+P^2}}{P}\,\,{\rm e}\,^{\pm\frac{i\pi
p}{\sqrt{\lambda }}}\;,\qquad P=\frac{\sqrt{\lambda }}{\pi
}\,\sin\frac{\pi p}{\sqrt{\lambda }}\;.
\end{equation}
The amplitude for $ZY\rightarrow ZY$ scattering\footnote{The
relevant component of the S-matrix is denoted by $L$ in
\cite{Beisert:2005tm,Klose:2006zd} and by $a_5$ in
\cite{Arutyunov:2006yd}.} is given by
\cite{Beisert:2005tm,Arutyunov:2006yd}
\begin{equation}\label{amp}
 S^{\rm string}=\frac{1-\frac{1}{x'_+x_-}}{1-\frac{1}{x'_-x_+}}\,\,
 \frac{x'_--x_+}{x'_+-x_-}\,
 \left(\frac{x'_--x_-}{x'_--x_+}\right)^2
 \,{\rm e}\,^{\frac{2i\pi p}{\sqrt{\lambda }}+i\theta (p,p')} \; ,
\end{equation}
where $x_\pm\equiv x_\pm(p_1)$, $x'_\pm\equiv x_\pm(p'_1)$.
The first term in the exponent indicates that we use the canonical
S-matrix in the string basis \cite{Arutyunov:2006yd} and the second term
is the dressing phase discussed in the introduction. It is a gauge dependent quantity, however, in the near-flat-space limit all generalized light-cone gauges become identical. For simplicity we choose therefore the uniform gauge, where the dressing phase is of the form \cite{Arutyunov:2004vx,Beisert:2005wv}:
\begin{equation}\label{chidef}
 \theta(p,p') = \frac{1}{\pi }\sum_{r,s=\pm} r s \,\chi(x_r,x'_s) \; ,
\end{equation}
with (we will only need this second derivative)
\begin{equation}\label{summa}
 \frac{\partial ^2\chi (x,y)}{\partial x\,\partial y}
 =\frac{\sqrt{\lambda }}{2}\sum_{r=2}^{\infty }\sum_{n=0}^{\infty }
 \frac{c_{r,n}}{x^ry^{r+2n+1}}-\left(x\leftrightarrow y\right)\;,
\end{equation}
and, to the one-loop accuracy \cite{Hernandez:2006tk},
\begin{equation}\label{}
 c_{r,n}=\delta _{n\,0}-\frac{8}{\sqrt{\lambda
 }}\,\,\frac{(r-1)(r+2n)}{(2r+2n-1)(2n+1)}+\ldots \; .
\end{equation}

We will now demonstrate that within the near-flat-space kinematics, the exact amplitude (\ref{amp}) agrees with (\ref{S-1loop}) upon expansion in $\pi /\sqrt{\lambda }$ and identification (\ref{glambda}). Let us first expand the phase $\theta (p,p')$. Taking into account that the difference between $P$ in (\ref{xpm}) and $p_1$ is unimportant at the one-loop level, we find from (\ref{chidef}), (\ref{xpm}):
\begin{equation}\label{}
 \theta (p,p')=-\frac{4\pi }{\lambda}\left(1+p_0\right)
 \left(1+p'_0\right)\left.\frac{\partial^2\chi }{\partial x\,\partial y}
 \right|_{x=\frac{1+p_0}{p_1}\,,\,\,y=\frac{1+p'_0}{p'_1}} \; .
\end{equation}
The summation in (\ref{summa}) yields
\begin{eqnarray}\label{}
 \frac{\partial^2\chi }{\partial x\,\partial y}&=&
 \frac{\sqrt{\lambda }}{2}\,\,\frac{x-y}{x^2y^2\left(xy-1\right)}
 \nonumber \\
 &&+\frac{2}{(xy-1)(x-y)}
 +\left[\frac{1}{(xy-1)^2}+\frac{1}{(x-y)^2}\right]
 \ln\frac{(x+1)(y-1)}{(x-1)(y+1)}+\ldots \; ,
\end{eqnarray}
and we get
\begin{eqnarray}\label{dess}
 \theta (p,p')&=&\frac{\pi }{\sqrt{\lambda }}\,\,
 \frac{\left[(p'_0-1)p_1-(p_0-1)p'_1\right]^2}{p'_0p_1-p_0p'_1}
  +\frac{4\pi }{\lambda }\,\,\left(\frac{p_1p'_1}{p'_0p_1-p_0p'_1}
 \right)^2
 \left[p'_0p_1-p_0p'_1
 \vphantom{\frac{(1+p_0+p_1)(1+p'_0-p'_1)}
 {(1+p_0-p_1)(1+p'_0+p'_1)}}\right.\nonumber \\
 &&\left.
 -p\cdot p'\ln\frac{(1+p_0+p_1)(1+p'_0-p'_1)}
 {(1+p_0-p_1)(1+p'_0+p'_1)}\right]+\ldots \;.
\end{eqnarray}

The real part of the  amplitude (the imaginary part of the S-matrix
element) comes entirely from the dressing phase,  and in the limit
(\ref{nearflatlimit}) becomes
\begin{equation}\label{}
 \Im S^{\rm string}_{\rm 1-loop}=
 \frac{8\pi }{\lambda }\,\,\frac{p_-^3p'{}^3_-}{p'{}^2_--p_-^2}
 \left(1-\frac{p'{}^2_-+p_-^2}{p'{}^2_--p_-^2}\,\ln\frac{p'_-}{p_-}\right) \; ,
\end{equation}
in complete agreement with (\ref{S-1loop}).

The  rest of (\ref{amp}), including the interference of the
tree-level phases, determines the absorptive part of the amplitude:
\begin{equation}\label{}
 \Re S^{\rm string}_{\rm 1-loop}
 =-\frac{\pi^2 }{2\lambda }\left[
 (p'_0p_1-p_0p'_1)^2+2(p_1^2-p_1p'_1-p'_1{}^2)
 +\frac{(p_1+p'_1)^2(p_1^2+p'_1{}^2)}{(p'_0p_1-p_0p'_1)^2}
 \right] \; .
\end{equation}
In the limit (\ref{nearflatlimit}) this becomes
\begin{equation}\label{}
 \Re S^{\rm string}_{\rm 1-loop}
 =-\frac{2\pi ^2}{\lambda }\,\frac{p_-^2p'{}^2_-\left(p'{}^2_-+p_-^2\right)}
 {\left(p'_--p_-\right)^2}\;,
\end{equation}
also in agreement with (\ref{S-1loop}).

The absorptive part of the one-loop amplitude can be reconstructed
from tree-level amplitudes by unitarity. Writing\footnote{Here we
switch from the $SO(4)^2$ notations in (\ref{main}) to the $SU(2)^4$
notations: $i'\rightarrow (a\dot{a})$, $i\rightarrow (\alpha
\dot{\alpha })$, see \cite{Klose:2006zd} for more details.}
$$
\Smatrix=\unit+\frac{2\pi i}{\sqrt{\lambda }}\,T,
$$
\begin{equation}
T \ket{Z_{\alpha\dot{\alpha}} Y'_{a\dot{a}}} =
   2L(p,p') \ket{Z_{\alpha\dot{\alpha}} Y'_{a\dot{a}}}
  - H(p,p') \ket{\Psi_{a\dot{\alpha}} \Upsilon'_{\alpha\dot{a}}}
  + H(p,p') \ket{\Upsilon_{\alpha\dot{a}} \Psi'_{a\dot{\alpha}}}
\end{equation}
and taking the tree-level amplitudes from
\cite{Klose:2006zd}\footnote{Here $a$ is a gauge parameter. The
gauge dependence disappears in  the limit (\ref{nearflatlimit}).}
\begin{align}
L(p,p') & = \frac{1}{4} \biggl[ (1-2a) \left(p'_0 p_1 - p_0 p'_1\right)
+ \frac{p_1^2-p'_1{}^2}{p'_0 p_1 - p_0 p'_1} \biggl] \; , \\
H(p,p') & = \frac{1}{2} \, \frac{p_1p'_1}{p'_0 p_1 - p_0 p'_1} \,
\frac{\left(p_0+1\right)\left(p'_0+1\right)-p_1p'_1}
{\sqrt{\left(p_0 +1\right)\left(p'_0+1\right)}} \; ,
\end{align}
we can use the optical theorem
\begin{equation}\label{}
 \Im T=\frac{\pi }{\sqrt{\lambda }}\,T^\dagger T
\end{equation}
to find the imaginary part of the one-loop contribution to
$S^{\rm string}$. In the limit of large $p_-$ we find
\begin{equation}
\bigabs{2 L(p,p')}^2 + 2 \bigabs{H(p,p')}^2 =
\frac{p_-^2p'{}^2_-\left(p'_-{}^2+p_-^2\right)}
 {\left(p'_--p_-\right)^2} \; ,
\end{equation}
which, multiplied by $\frac{\pi}{\sqrt{\lambda}}\cdot\frac{2\pi}{\sqrt{\lambda}}$, is exactly what we have obtained before.

\section{Conclusions and outlook}

The string sigma-model on $AdS_5\times S^5$ simplifies considerably
in the near-flat space limit thus making loop computations feasible.
This opens up a possibility to check various conjectures about the
exact S-matrix or the spectrum of the AdS string. It is not obvious
that the reduced theory agrees with the full sigma-model at the
quantum level, because low-momentum states could survive in loop
diagrams even if the external legs all have large light-cone
momenta. For instance, the momentum flowing through the t-channel loop in diagram fig.~\ref{scatter}~(e) is zero. However, the agreement of the one-loop scattering amplitudes strongly suggests that the low-momentum states indeed decouple.

Another indication of the self-consistency of the near-flat space
reduction is the finiteness of the one-loop amplitudes. The
divergences cancel due to the asymmetric treatment of left- and right-moving modes. We believe that the same mechanism renders the model finite to all loop orders.

\bigskip
\subsection*{Acknowledgments}
\bigskip
We would like to thank T.~McLoughlin, R.~Roiban and especially
J.~Minahan for interesting discussions. We would also like to thank
J.~Minahan for collaboration on the early stages of this project.
The work of K.Z. was supported in part by the Swedish Research
Council under contract 621-2004-3178, by grant NSh-8065.2006.2 for
the support of scientific schools, and by RFBR grant 06-02-17383.
The work of T.K. and K.Z. was supported by the G\"oran Gustafsson
Foundation.

\bibliographystyle{nb}
\bibliography{nearflat}

\end{document}